\documentclass[11pt,a4paper,final]{article}
\usepackage{amsmath}
\usepackage{amsfonts}
\usepackage{amssymb}
\usepackage{graphicx}
\usepackage{algorithm} 
\usepackage{algpseudocode}
\usepackage{natbib}
\usepackage[margin=1.2in]{geometry}
\usepackage{color}
\usepackage{bm}
\usepackage{hyperref}
\usepackage{placeins}
\usepackage{amsthm}
\usepackage{authblk}
\usepackage{caption,subcaption}
\usepackage{ulem}

\begin{document}

\title{Comments on ``Bayesian Solution Uncertainty Quantification for Differential Equations" by Chkrebtii, Campbell, Calderhead \& Girolami}

\author[1,2]{Fran\c{c}ois-Xavier Briol}
\author[1]{Jon Cockayne} 
\author[2]{Onur Teymur}

\affil[1]{Department of Statistics, University of Warwick}
\affil[2]{Department of Mathematics, Imperial College London}

\maketitle

\abstract{We commend the authors for an exciting paper which provides a strong contribution to the emerging field of probabilistic numerics (PN). Below, we discuss aspects of prior modelling which need to be considered thoroughly in future work.}

\subsubsection*{Introduction}

The majority of PN solvers (including the present paper) take a Bayesian viewpoint and hence require several modelling choices including prior specification. As with any inference problem, there exists a trade-off between representing prior beliefs and choosing a prior which is convenient and/or readily interpretable mathematically. We believe that the consequences of these assumptions are not often discussed in enough detail and therefore highlight below several issues.

\paragraph{Computational Complexity}

Of interest was the discussion into reduction of the computational complexity by exploiting compactly supported covariance function. The authors note in Sec.~3.2 that while such a choice will yield a method involving inversion of a sparse matrix, this is not explored further -- though this will have an effect on the rate of convergence of the estimator. We believe that a study of the extent of this effect is of some importance, as there is a clear trade-off here between steps desired to achieve a required tolerance, and the computational cost of each step.

\paragraph{Tractability}

One issue is the intractability of the joint conditional predictive probability distribution in Sec.~2.1 which depends on analytically convolving covariance functions. This is not possible except for a few simple kernels; relying upon such construction therefore significantly restricts the range of priors available. 

With this in mind, we note that differentiating kernels is often easier than integrating them. Unless there is a specific reason to model $u_t$, it may therefore be more convenient to define a kernel for $u$ and differentiate it to obtain a kernel for $u_t$.  

Another interesting point is that this trade-off is also encountered in Bayesian Quadrature, a PN method for integration. A table of kernels which can be integrated analytically is provided in \cite{Briol2016} and may be of interest to users of the present methodology.

\paragraph{Boundary conditions}

An important point for PDEs is how best to make use of boundary information. The authors observe that, for ODEs, it is simple to encode the initial condition in the prior, but generalising this to PDEs is significantly more challenging owing to the fact that the boundaries will now typically be a manifold of dimension larger than zero. 

Significant work in this area includes that of \cite{Owhadi2015} and \cite{Cockayne2016} which select covariances based on Green's functions, though the computations involved are challenging and such closed-form conditioning is narrowly applicable as a result. In general \cite{Owhadi2015Conditioning} shows that conditioning over the entire boundary is well-defined from a mathematical perspective, provided the boundary operator is linear. We would be interested in whether this can be generalised in a tractable way so that, for example, we can define a prior over those functions which satisfy the boundary conditions exactly.

\paragraph{Relationship to known integrators}

A desideratum (although not always a requirement) for a probabilistic method is that the estimate given by some readily-calculated statistic of the posterior distribution corresponds to the output of a classical numerical solver. The advantage here is that the theory of such solvers is highly developed and certain properties -- convergence, stability, etc. -- can potentially be inherited. This method does not, to the best of our knowledge, satisfy this property. However, more recent work which builds on this work includes a general construction provided by \cite{Conrad2015} and careful choice of the kernel within a similar framework has subsequently been shown to correspond to Runge-Kutta methods of order less than four \citep{Schober2014} and linear multi-step methods of arbitrary order \citep{Teymur2016}.

\subsubsection*{Conclusion}

Once again, we would like to congratulate the authors of this paper which provided foundational work upon which many subsequent PN methods have been built. We hope to have highlighted some of the important issues relating to the development of priors which will hopefully influence future work in this area.

\paragraph{Acknowledgements} This work was completed within the Probabilistic Numerics working group which is part of the 2016-2017 SAMSI programme on Optimization. FXB was also supported by EPSRC [EP/L016710/1].

{ \scriptsize
\bibliographystyle{plainnat}
\bibliography{discussion_bib}

}

\end{document}